\documentclass[conference]{IEEEtran}
\ifCLASSINFOpdf
\else
\fi
\usepackage[letterpaper, left=1in, right=1in, bottom=1in, top=0.75in]{geometry}
\usepackage[T1]{fontenc}% optional T1 font encoding
\usepackage[linesnumbered,ruled]{algorithm2e}
\usepackage{stfloats,color}
\usepackage{amsfonts}
\usepackage{amssymb}
\usepackage[cmex10]{amsmath}
\usepackage{graphicx}
\usepackage{setspace}
\usepackage{subcaption}
\usepackage{cite}
\usepackage{array}
\usepackage{mdwmath}
\usepackage{mdwtab}
\usepackage{xcolor}
\usepackage{times}
\usepackage{epsfig}
\usepackage{latexsym}
\usepackage{epstopdf}
\usepackage{verbatim}
\usepackage{units}
\usepackage{amsthm}
\usepackage{mdwlist}
\usepackage{placeins}
\usepackage{afterpage}
\usepackage{dsfont}
\usepackage{soul}
\DeclareMathOperator*{\argmin}{\arg\!\min}
\DeclareMathOperator*{\argmax}{\arg\!\max}
\newcommand{\defeq}{\ensuremath{\triangleq}}
\usepackage[colorinlistoftodos]{todonotes}

\usepackage{todonotes}

\newtheorem{definition}{Definition}
\newtheorem{lemma}{Lemma}

\makeatletter
\def\blfootnote{\xdef\@thefnmark{}\@footnotetext}
\makeatother

\begin{document}

\title{Delay-Aware Coded Caching for Mobile Users }

\author{\IEEEauthorblockN{Emre Ozfatura$^{*}$, Thomas Rarris$^{*}$, Deniz G{\"u}nd{\"u}z$^{*}$, and Ozgur Ercetin$^{\dagger}$}
\IEEEauthorblockA{$^{*}$Information Processing and Communications Lab\\
Department of Electrical and Electronic Engineering, Imperial College London\\
\{m.ozfatura, thomas.rarris14, d.gunduz\}@imperial.ac.uk
\\$^{\dagger}$Sabanci University, Turkey, oercetin@sabanciuniv.edu}
}


\maketitle

\begin{abstract}
In this work, we study the trade-off between the cache capacity and the user delay for a cooperative Small Base Station (SBS) coded caching system with mobile users. First, a delay-aware coded caching policy, which takes into account the popularity of the files and the maximum re-buffering delay to minimize the average re-buffering delay of a mobile user under a given cache capacity constraint is introduced. Subsequently, we address a scenario where some files are served by the macro-cell base station (MBS) when the cache capacity of the SBSs is not sufficient to store all the files in the library.  For this scenario, we develop a coded caching policy that minimizes the average amount of data served by the MBS under an average re-buffering delay constraint.
\end{abstract}
\blfootnote{This work was supported in part by the Marie Sklodowska-Curie Actions SCAVENGE (grant agreement no. 675891) and  TACTILENet (grant agreement no. 690893), and by the European Research Council (ERC) Starting Grant BEACON (grant agreement no. 725731).}

% \fi
\IEEEpeerreviewmaketitle
\section{Introduction}
During last decade, the on-demand video streaming applications have been dominating the bulk of the Internet traffic. In 2016, YouTube alone was responsible for 21\% of the mobile Internet traffic in North America \cite{sandvine}. According to Cisco Visual Networking Index report \cite{cisco}, the size of the Internet video traffic will be four times larger by the year 2021. This rapid increase in the Internet video traffic calls for a paradigm shift in the design of cellular networks. A recent trend is to store the popular content at  the  network  edge, closer to the user, in order to mitigate the excessive video traffic in the backbone.\\
\indent In heterogeneous cellular networks, SBSs can be equipped with storage devices, containing popular video files, to reduce the latency as well as the transmission cost. In a network of densely deployed SBSs, there may be more than one SBS that can serve the requested content of a mobile user (MU). This flexibility in user assignment is exploited in designing cooperative caching policies \cite{SC.coopR1,SC.coopR2,SC.coopR3}, wherein the main objective is  to minimize the transmission cost of serving user requests. It has also been shown that storing the contents in a coded form, particularly using maximum distance separable (MDS) codes, utilizes the local storage more efficiently; thereby increasing the amount of data served locally \cite{SC.CoD1,SC.CoD2}.\\ 
\indent However, aforementioned works seek to find an optimal cooperative caching policy based on a given static user access topology such that the closest SBS to a user do not change over the time. However, in ultra dense networks (UDNs), due to the limited coverage area of SBSs, user access patterns are usually dynamic, and the mobility patterns of users have a significant impact on the amount of data that can be delivered locally\cite{SC.M2}. To this end, mobility-aware cooperative caching policies have been recently studied  in \cite{SC.M3,SC.M1}. In these works, the goal is to maximize the amount of data that is served locally while satisfying a given content downloading delay constraint. However, when the contents are stored in  a coded form as in \cite{SC.M3,SC.M1}, a user cannot start displaying the video content before collecting all the parity bits, which may cause significant initial buffering delays in video streaming applications.\\ 
\indent Proactive content caching for the continuous video display scenario, in which users can start displaying video content before downloading all the video fragments has been previously studied in \cite{SC.Mdn1} where, SBSs fetch the content dynamically in advance, prior to user arrivals, using the instantaneous user mobility information. Instead of a dynamic content fetching policy, in this paper, we consider a static caching policy similarly to \cite{SC.M3} and \cite{SC.M1}, and focus on the continuous display of video. 
\section{System Model and Problem Formulation}
Consider a heterogeneous cellular network that consists of one MBS and $N$ SBSs, denoted by ${\mathrm{SBS_{1}},\ldots,\mathrm{SBS_{N}}}$, with disjoint coverage areas of the same size. Further, each SBS is equipped with a cache memory of size $C$ bits. Due to disjoint coverage, a MU is served by only one SBS at any particular time. We assume that time is divided into equal-length time slots, and the duration of a time slot corresponds to the minimum time that a MU remains in the coverage area of the same SBS. We also assume that each SBS is capable of transmitting $B$ bits to a MU within its coverage area in a single time slot.\\ 
\indent For user requests, we consider library of $K$ video files $\mathbb{V} = \{v_1, \ldots, v_K\}$, each of size $F$ bits. Video files in the library are indexed according to their popularity, such that $v_{k}$ is the $k$th most popular video file with a request probability of $p_{k}$. Since the size of a video file is $F$ bits and the transmission rate of a SBS is $B$  bits per time slot, a MU can download a single video file in at least $T=F/B$ time slots. For the sake of simplicity, we assume that $T$ is an integer, and we call the $T$ time slots following a request as a \textit{video downloading session}. Although, a MU is connected to only one SBS at each time slot, due to mobility, it may connect to multiple SBSs within a video downloading session. Due to the limited cache memory size, all video files in the library may not be stored at SBSs and in that case requests for the uncached video files are offloaded to the MBS. 
\subsection{User Mobility}
%%%%%%%%%%%%%%%%%
\begin{figure}
    \centering
          \includegraphics[scale=0.3]{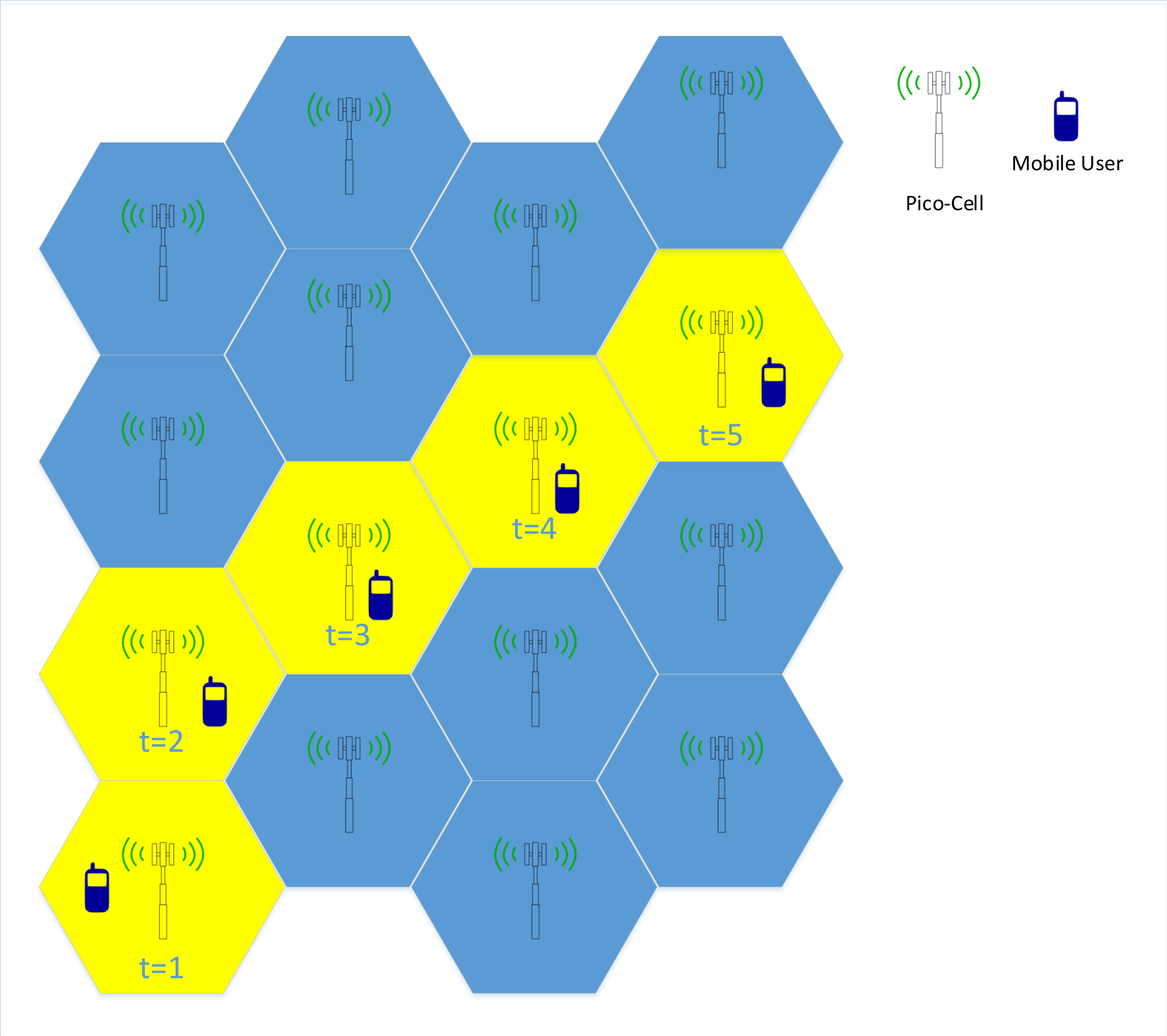}
  				\caption{A sample mobility path under the high mobility assumption for $T=5$.}\label{f:mobility_path}
 \end{figure}
%%%%%%%%%%%%%%%%%

 {\em Mobility path} of a MU is defined as the sequence of SBSs visited within a video downloading session. For instance, for $T=5$, $SBS_{1},SBS_{3},SBS_{4},SBS_{5},SBS_{6}$ is a possible mobility path. We consider a {\em high mobility} scenario, in which a MU does not stay connected to the same SBS more than one time slot so that at the end  each time slot, MU moves to one of the  neighboring cells as illustrated in Figure \ref{f:mobility_path}. Under this assumption, a mobility path is a sequence of $T$ distinct SBSs. 
\subsection{Delay-aware coded caching}
\begin{figure}
    \centering
        \includegraphics[scale=0.35]{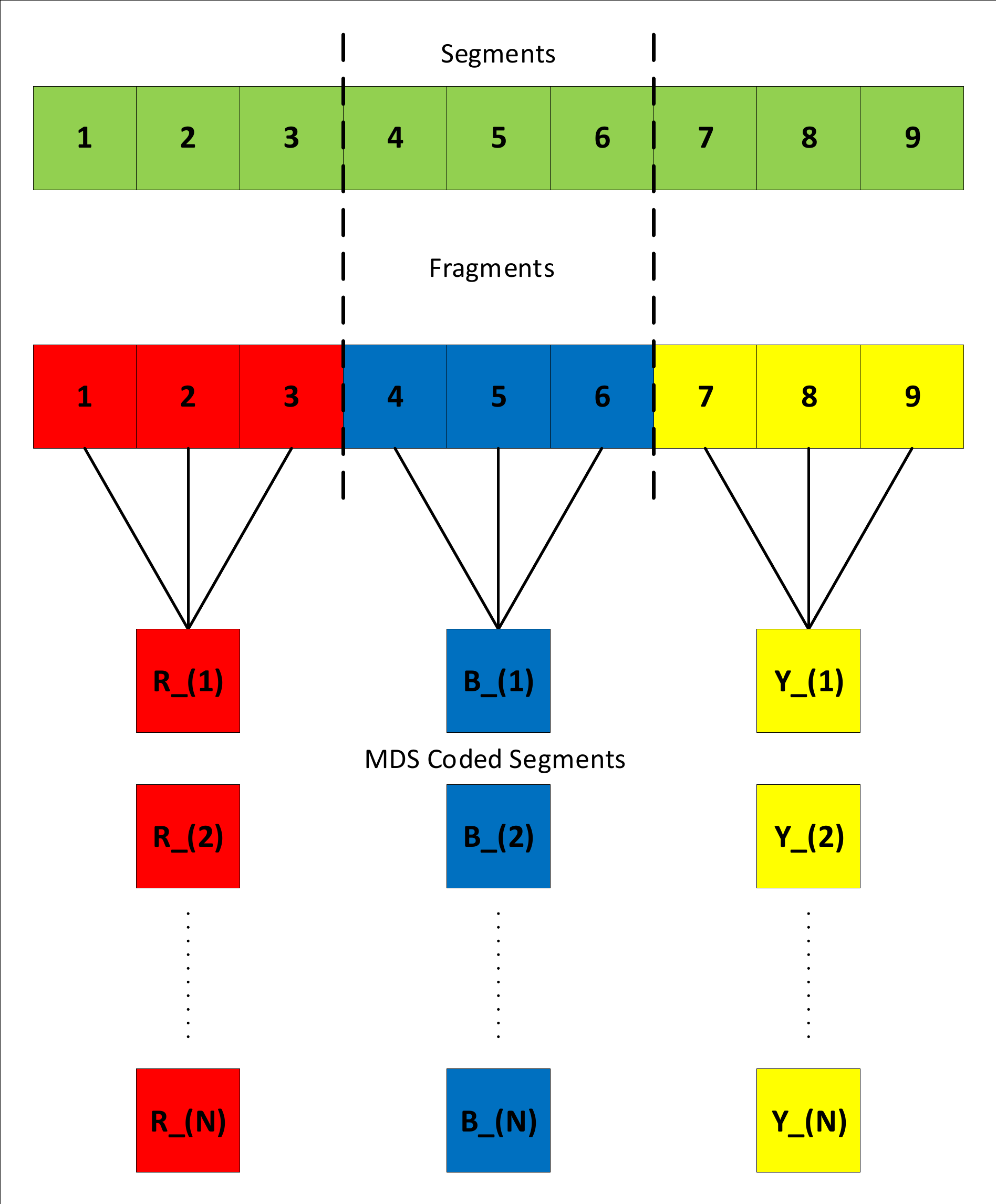}
				\caption{Illustration of the employed coded caching strategy for a video download session of $T=9$ time slots.}\label{f:coding_strategy}
\end{figure}

Before proceeding with the problem formulation, we explain the coding scheme that is used to encode the video files. First, each video file is divided into $T$ disjoint video segments of size  $B$ bits each, i.e., 
$v_{k}=\left(s^{(1)}_{k},\ldots,s^{(T)}_{k}\right)$. Second, these segments are grouped into $M_{k}$ disjoint fragments $f^{(1)}_{k}, \ldots, f^{(M_{k})}_{k}$; that is,
\begin{equation}
v_{k}=\bigcup^{M_{k}}_{m=1}f^{(m)}_{k}, \text{~~and } f^{(i)}_{k}\cap f^{(j)}_{k}=\emptyset,
\end{equation}
for any $i,j\in \left\{ 1,\ldots,M_{k}\right\}$ and $i \neq j$. Then, the segments in each fragment are jointly encoded using a $\left(\lvert f^{(m)}_{k}\rvert,N\right)$ MDS code, and each coded segment is cached by a different SBS. Hence, any fragment $f^{(m)}_{k}$  can be recovered from any $\lvert f^{(m)}_{k}\rvert B$ parity bits collected  from any $\lvert f^{(m)}_{k}\rvert$ different SBSs within $\lvert f^{(m)}_{k}\rvert$ time slots.\\
\indent The video encoding strategy is illustrated by the following example for $T=F/B=9$. A video file is first divided into $T=9$ segments, which are then grouped into three {\em fragments} of  three segments each (each fragment is represented by a different color in Figure 2). The three segments in each fragment are jointly encoded using a (3,N) MDS code to obtain $N$ different coded segments of size $B$ bits each. Then, each coded segment is cached by a different SBS, e.g., the $i$th coded segment of each file is cached by $SBS_{i}$. The overall coded caching procedure is illustrated in Figure \ref{f:coding_strategy}. The reason for constructing $N$ coded segments is to ensure that in any possible path a MU does not receive the same coded segment multiple times. We remark that for given $T$, certain cells can not be visited in a same mobility path, hence, depending on $T$, less than $N$ coded segments might be sufficient to prevent multiple reception of the same coded segment. \cite{WSA}. 
\begin{definition}
A coded caching policy $\mathbf{X}$ defines how each file $v_{k}$ is divided into fragments, i.e., $\mathbf{X}\defeq \left\{\mathbf{X}_{k}\right\}^{K}_{k=1}$, where $\mathbf{X}_{k}=\left\{f^{(1)}_{k},\ldots,f^{(M_{k})}_{k}\right\}$.
\end{definition}
Note that since the cache capacity of a SBS is $C$ bits and the size of each coded segment is $B$ bits, a feasible caching policy should satisfy the inequality
$\sum^{K}_{k=1}M_{k}B\leq C$.

%%%%%%%%%%%%%%%%%
%%%%%%%%%%%%%%%%%

\subsection{Continuous video display and delay analysis}
The \textit{video display rate}, $\lambda$, defines the average amount of data (bits) required to display a unit duration (normalized to one time slot) of a video file\footnote{In general, video files are variable bit rate (VBR) encoded, and the display rate varies over time. However, $\lambda$ can be considered as the minimum value satisfying $\lambda t \geq \lambda_{c}(t)$, where $\lambda_{c}(t)$ is the cumulative display rate of a VBR-encoded video. Hence, the delay requirements can be satisfied at a constant rate of $\lambda$.}. In this work, we consider the scenario in which the service rate of the SBSs and the video display rate of MUs are approximately equal, i.e., $ B\approx \lambda$. Hence, at each time slot 
a MU displays one segment and similarly downloads one coded segment. In order to display a segment, it should be available at the buffer in an uncoded form. If the corresponding segment is not available in the buffer, then the user waits until the corresponding segment is available at the buffer. This waiting time is called the \textit{re-buffering delay}.\\
\indent The cumulative re-buffering delay 
for file $v_{k}$, under policy $\mathbf{X}_{k}$, is denoted by $D_{k}(\mathbf{X}_{k})$, and it is equal to the sum of re-buffering delays experienced within a video streaming session. For the delay analysis, lets consider a particular file which is divided into $M$ fragments, i.e., $\left\{f^{(1)},\ldots,f^{(M)}\right\}$. The \textit{display duration} of a fragment is the number of segments in it, e.g., if there is only one fragment then the display duration of that fragment is equal to the video duration. Let $d^{(m)}$ denote the display duration of the $m$th fragment, i.e.,  $d^{(m)}=\lvert f^{(m)}\rvert B/\lambda \approx \lvert f^{(m)}\rvert$~. Furthermore, let $t^{(m)}_{d}$ and $t^{(m)}_{p}$ denote the time instants at which the $m$th fragment is downloaded and started to be displayed, respectively. If  $t^{(m)}_{p}>t^{(m)}_{d}$, the user displays the $m$th fragment without experiencing a stalling event; however, if $t^{(m)}_{d}>t^{(m)}_{p}$, then the user enters a re-buffering period and it stops displaying the video until  $t^{(m)}_{d}$. Accordingly, the re-buffering duration for the $m$th fragment, $\Delta^{(m)}$, can be formulated as
\begin{equation}\label{rebuffereq}
\Delta^{(m)}= \max \left\{ t^{(m)}_{d}-t^{(m)}_{p},0 \right\}.
\end{equation}
Note that $t^{(m)}_{p}$ is equivalent to the sum of the display times and re-buffering delays experienced by the previously displayed fragments, i.e.,
\begin{equation}
t^{(m)}_{p}=\sum^{m-1}_{i=1}\Delta^{(i)} + d^{(i)}.
\end{equation}
\indent Similarly, assuming that the fragments are downloaded in order, $t^{(m)}_{d}$ is the total download time of all the previous fragments, i.e.,
\begin{equation}
 t^{(m)}_{d}=\sum^{m}_{i=1}d^{(i)}.
\end{equation}
Hence, (\ref{rebuffereq}) can be rewritten as
\begin{equation}
\Delta^{(m)} = \max \left\{d^{(m)}-\sum^{m-1}_{i=1}\Delta^{(i)},0 \right\}.
\end{equation}
We observe that if $\Delta^{(m)}>0$, then the following equality holds,
\begin{equation}\label{delta-d} 
\sum^{m}_{i=1}\Delta^{(i)}= d^{(m)}. 
\end{equation} 
Let $D$ be the cumulative re-buffering delay experienced over all fragments of the video, which is derived by the following lemma.
\begin{lemma}
Cumulative re-buffering delay $D$ is equal to the display duration of the largest fragment, i.e.,
\begin{equation}\label{delaymodel}
D=\sum^{M}_{m=1}\Delta^{(m)}=\max\left\{d^{(1)},\ldots,d^{(m)}\right\}.
\end{equation}
\end{lemma}
Lemma 1 can be easily proved by induction using equality (\ref{delta-d})  and the fact that $\Delta^{(1)}=d^{(1)}$. Note that if the first fragment has the largest display duration, then $D=\Delta^{(1)}$ and the cumulative re-buffering delay is equal to the initial buffering delay. 
\subsection{Problem formulation}
In this work, we aim to find the optimal coded data caching policy $\mathbf{X}$ that minimizes the cumulative re-buffering delay averaged over all files, i.e., $D_{avg}(\mathbf{X})=\sum^{K}_{k=1}p_{k}D_{k}(\mathbf{X}_{k})$. Before presenting the problem formulation, we focus on a particular file and highlight the delay-cache capacity trade-off with an example. If the number of fragments is equal to the number of segments, i.e., $f^{(m)}=\left\{s^{(m)}\right\}$ $\forall m\in \left\{1,\ldots,T\right\}$, then each SBS caches all the segments. This requires a memory of $F=TB$ bits for the corresponding file. On the other hand, if there is only one fragment that contains all the segments, i.e., $f^{(1)}=\left\{s^{(1)},\ldots,s^{(T)}\right\}$, then all the segments are jointly encoded, and each SBS caches only $B$ bits for the corresponding file. Note that, although the download time of the content is $T$ slots in both cases; in the first case, each fragment can be displayed right after downloading it; whereas, in the second case, it is not possible to start displaying a fragment until all the $F=TB$ parity bits are collected, since all the segments are encoded jointly. Equivalently, the cumulative re-buffering delay is equal to one time slot in the first case and $T$ slots in the second.\\
%%%%%%%%%%%%%%%%%%
\begin{figure}
  \centering
  \includegraphics[scale=0.55]{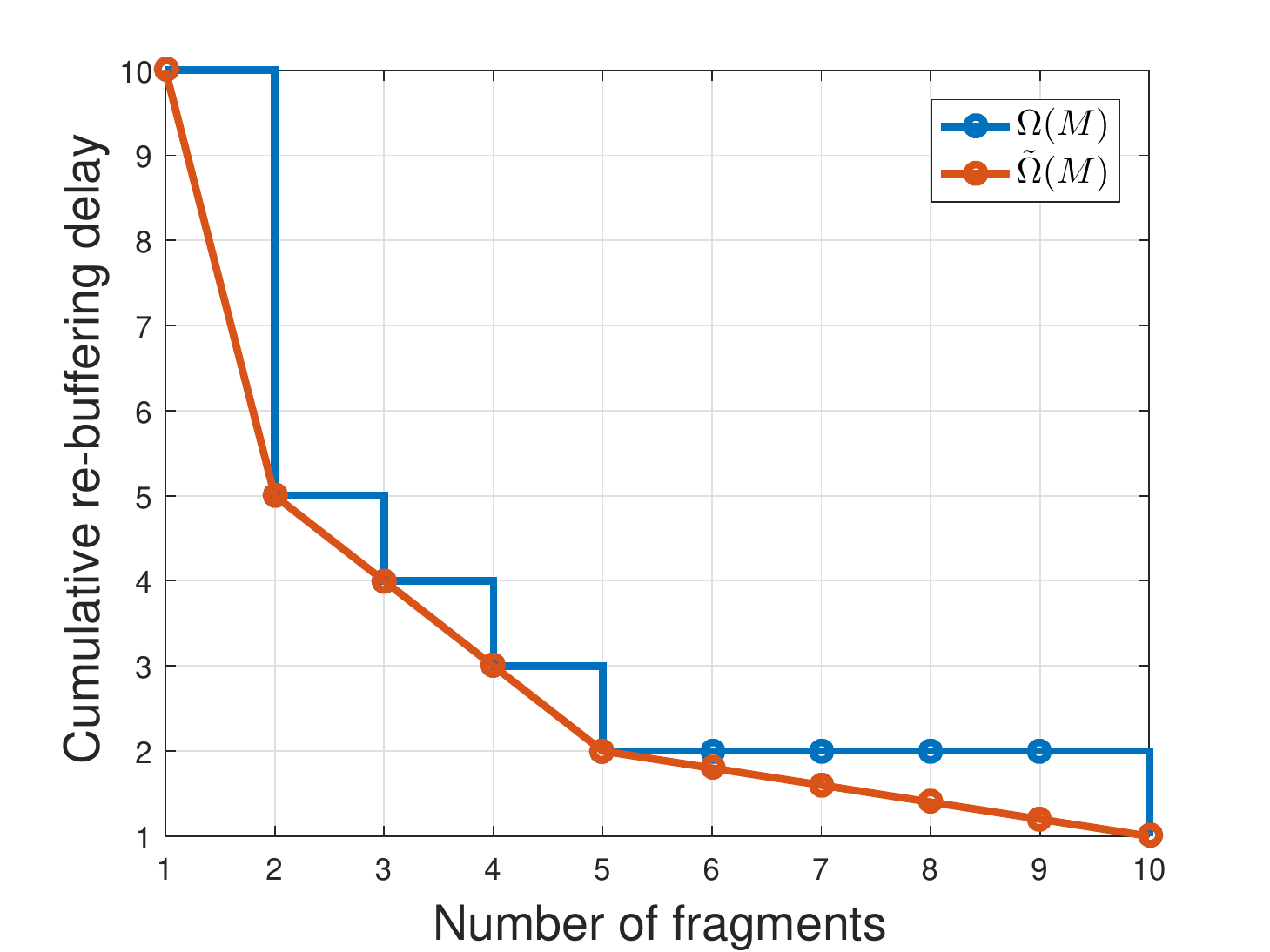}
  \caption{Delay-cache capacity function and its piece-wise linear approximation for $T=10$}
  \label{fig:delayLevels}
\end{figure}
%%%%%%%%%%%%%%%%%%
\indent Next, we introduce a general mathematical model for the delay-cache capacity trade-off. The required cache size for a file depends only on the number of fragments $M$, and it is $MB$ bits. However, the cumulative re-buffering delay is equal to the display time (the number of segments) of the largest fragment. Hence, for given $M$ the cumulative re-buffering delay can be minimized by choosing fragment sizes approximately equal, i.e., for any $i,j\in \left\{1,\ldots,M\right\}$ and $i\neq j$, $ \lvert d^{(i)}-d^{(j)} \rvert \leq 1$. Consequently, for a given memory constraint of $MB$ bits the minimum achievable cumulative re-buffering delay is equal to $\left\lceil T/M\right\rceil$ time slots.\\
\indent To mathematically capture this relationship, we introduce the {\em delay-cache capacity} function $\Omega(M)\defeq\left\lceil T/M\right\rceil$ which maps the number of fragments in a file to the minimum achievable re-buffering delay $D$. $\Omega(M)$ is a monotonically decreasing step function which is illustrated in Figure \ref{fig:delayLevels} for $T=10$. To analyze $\Omega(M)$, we introduce two new parameters: the \textit{delay level} and the \textit{decrement point}. Any possible value of $\Omega(M)$ is called delay level and denoted by $D^{(l)}$. For the given example illustrated in Figure \ref{fig:delayLevels}, there are $L=6$ delay levels,  i.e., $D^{(1)}=10$, $D^{(2)}=5$, $D^{(3)}=4$, $D^{(4)}=3$, $D^{(5)}=2$, $D^{(6)}=1$. A decrement point $m^{(l)}$ is the minimum value of $M$ that satisfies $\Omega(M)=D^{(l)}$. In the given example $m^{(1)}=1$, $m^{(2)}=2$, $m^{(3)}=3$, $m^{(4)}=4$, $m^{(5)}=5$, $m^{(6)}=10$.\\
\indent Recall that popularity of the files are not identical, which implies that re-buffering delay of the popular files has more impact on the average re-buffering delay. Hence, for each file $v_{k}$, we consider a weighted delay-cache capacity function $\Omega_{k}(M_{k})$ such that $\Omega_{k}(M_{k})\defeq p_{k}\left\lceil T/M_{k}\right\rceil$. Note that for a given number of fragments $M$, we know the optimal caching decision, i.e., the number of segments in each fragment. Hence, from now on, we use  $\mathbf{M}\defeq(M_{1},\ldots,M_{K})$ to denote the caching policy instead of $\mathbf{X}$. Then the average re-buffering delay is rewritten as $D_{avg}(\mathbf{M})=\sum^{K}_{k=1}\Omega_{k}(M_{k})$. Eventually, we have the following optimization problem
\begin{small}
\begin{align}
   \text{\bf P1:}~~~~~~~ \;\;\;\min_{\mathbf{M}}&
   \begin{aligned}[t]
      D_{avg}(\mathbf{M})\notag
   \end{aligned}\\
   \text{subject to: } &D_{k}(M_{k})\leq D_{max},~ \forall k, \label{fairconst}\\
	 &\sum^{K}_{k=1}M_{K}B\leq C, \label{memory}
\end{align}
\end{small}            
where (\ref{fairconst}) is the fairness constraint which ensures that the cumulative re-buffering delay is less than $D_{max}$ for any video file, and (\ref{memory}) is the cache capacity constraint.
\section{Solution Approach}
Lets denote the minimum $l$ that satisfies 
$D^{(l)}<D_{\max}$ in {\bf P1} by $l_{\min}$. Then, the optimization problem  \text{\bf P1} can be reformulated as
\begin{small}
\begin{align}
   \text{\bf P2:}~~~~~~~~ \;\;\;\min_{\mathbf{M}}&
   \begin{aligned}[t]
      D_{avg}(\mathbf{M})  \notag
   \end{aligned}\\
   \text{subject to: } &M_{k}\geq m^{(l_{\min})}, \forall k, \label{P2_cond1}\\
	 &  \sum ^{K}_{k=1} M_{k}\leq C/B.  
\end{align}
\end{small}
Note that we simply converted the delay constraint to a cache capacity constraint, such that each file requires a cache capacity of at least $m^{(l_{\min})}B$ bits. In order to find a feasible solution to $\text{\bf P2}$ the cache capacity $C$ should be larger than $Km^{(l_{\min})}B$ bits. In the following section, we first solve \text{\bf P2} assuming that this condition holds. We will consider the other case in the subsequent section. Note that, if (\ref{P2_cond1}) does not hold for all files, then  some of the least popular files are not cached at all, and a MU requesting one of these files is offloaded to the MBS causing additional overhead. Later we will show how this overhead is modeled. We define a caching strategy as {\em Cost-free} if all
the video files are cached by SBSs.
\subsection{Cost-free delay minimization}
\text{\bf P2} can be shown to be an NP hard problem, since it can be reduced to a knapsack problem. However, if we use a piecewise linear approximation of the delay-cache capacity function $\Omega_{k}(M_{K})$, which is denoted by $\tilde{\Omega}_{k}(M_{K})$, then the objective function becomes the sum of piecewise monotonic linear functions. Let $\gamma_{k,l}$ be the slope of the function $\tilde{\Omega}_{k}(M_{K})$, in the interval $(m^{(l)} m^{(l+1)}]$. Then, it is easy to observe that $\vert\gamma_{k,l}\vert>\vert\gamma_{k,l+1}\vert$ holds for all $l$. Hence, if the objective function is replaced by $\tilde{D}_{avg}(\mathbf{M})=\sum^{K}_{k=1}\tilde{\Omega}_{k}(M_{K})$, we obtain the following convex optimization problem: 
 \begin{small}
\begin{align}
   \text{\bf P3:}~~~~~~~ \;\;\;\min_{\mathbf{M}}&
   \begin{aligned}[t]
      \tilde{D}_{avg}(\mathbf{M})=\sum^{K}_{k=1}\tilde{\Omega}_{k}(M_{K})  \notag
   \end{aligned}\\
   \text{subject to: }  &M_{k}\geq m^{(l_{\min})} \text{ for all } k\\ \label{sizeconst}
	 &  \sum ^{K}_{k=1} M_{k}\leq C/B.  
\end{align}
\end{small}
\begin{algorithm}[t]{\footnotesize
    \SetKwInOut{Input}{Input}
    \SetKwInOut{Output}{Output}
    \Input{$B$,$C$,$\left\{\left\{\gamma_{k,l}\right\}^{L}_{l=1}\right\}^{K}_{k=1}$}
    \Output{$\mathbf{M}$}
		    $M_{k} \gets m^{(l_{\min})},k\in\left\{1,\ldots,K\right\}$\;
				$\gamma_{k}\gets \gamma_{k,l_{\min}},k\in\left\{1,\ldots,K\right\}$\;
				$l_{k}\gets l_{\min}$\;
				$\tilde{C} \gets C/B$\;
        \While{$\tilde{C}>0$} {
		     $\acute{k}=\argmax\left\{\gamma_{1},\ldots,\gamma_{K}\right\}$\;
				 \eIf{$\tilde{C}\geq (m^{(l_{\acute{k}}+1)}-m^{(l_{\acute{k}})})$}{
                 $l_{\acute{k}}\gets l_{\acute{k}}+1$\;
								$\gamma_{\acute{k}}\gets \gamma_{k,l_{\acute{k}}}$\;                
				 $M_{k} \gets m^{(l_{\acute{k}})}$\;                			
			    $C_{B}\gets C_{B}-(m^{(l_{\acute{k}})}-m^{(l_{\acute{k}}-1)})$
				      }{
		          $M_{\acute{k}} \gets M_{\acute{k}}+C_{B}$\;
							$\tilde{C}\gets 0$\;
							 }
													}
    \caption{Cost-free delay minimization}}
\end{algorithm}
Note that the solution of \text{\bf P3} is not equivalent to the solution of the original problem \text{\bf P2}. However, we will show that with a small perturbation in the cache size $C$, solution of \text{\bf P2} and \text{\bf P3} becomes identical. Since the objective is a convex function of sum of piecewise linear functions, we follow a similar strategy to the one used in \cite{SC.M1}. The proposed algorithm  first allocates each file a cache memory of size $m^{(l_{\min})}B$ bits, which corresponds to the delay level of $D^{(l_{\min})}$. After this initial phase, it searches for the $\tilde{\Omega}_{k}(M_{k})$ that has the minimum slope (maximum negative slope), and updates the delay level of file $v_{k}$ to the next one, i.e., $D^{(l)}$ to $D^{(l+1)}$, and updates $M_{k}$ accordingly. The procedure is repeated until (\ref{sizeconst}) is satisfied with equality. The overall coded caching strategy is detailed in Algorithm 1.
 \indent Note that $\Omega(m^{(l)})=\tilde{\Omega}(m^{(l)})$ at any decrement point $m^{(l)}$ by construction, as illustrated in Figure \ref{fig:delayLevels}. Hence, if for each $k$, equality $M_{k}=m^{(l_{k})}$ holds for some $l_{k}\in\left\{l_{min},\ldots,L\right\}$, then $D_{avg}(\mathbf{M})$ is equal to $\tilde{D}_{avg}(\mathbf{M})$. Equivalently, if Algorithm 1 terminates in \emph{if condition}, then for the resulting caching policy $M$, $D_{avg}(\mathbf{M})= \tilde{D}_{avg}(\mathbf{M})$. Now recall that by construction $\tilde{D}_{avg}(\mathbf{M})$ is a lower bound for $D_{avg}(\mathbf{M})$ which then implies that $\mathbf{M}$ is the optimal solution for the original problem  {\bf P2}. If Algorithm 1 terminates in \emph{else condition}, then the obtained policy will be a suboptimal solution for {\bf P2}. Nevertheless, it is always possible to ensure that last cache size allocation is done in \emph{if condition} via increasing cache size $C$ by $\epsilon\leq F/2$ since $m^{l+1}-m^{l}\leq BT/2=F/2$ for any $l$.
\subsection{Average delay constrained cost minimization}
In some cases, it may not
be possible to satisfy the $D_{max}$ constraint for all files in the library due to cache capacity constraints. Furthermore, the average re-buffering delay can be a predefined system parameter, denoted by $D_{avgMax}$, in order to offer a certain QoS to the user; however, the average delay obtained from the solution of {\bf P2} may not satisfy this requirement. As a result, some of the least popular files are not cached at all and the requests for these files are offloaded to MBS.\\
\indent We denote the average amount of data that needs to be downloaded from the MBS by $\Theta$ and let be the set of cached videos, $A = \{k : M_k > 0\}$ then $\Theta=\sum_{k \notin A}p_{k}$. Our goal is to find the coded caching policy $\mathbf{M}$ that minimizes $\Theta$, thus we have the following optimization problem:
\begin{small}
\begin{align}
   \text{\bf P4:}~~~~~~ \;\;\;&\min_{\mathbf{M}}
   \begin{aligned}[t]
      \Theta(\mathbf{M})=\sum_{k \notin A}p_k \notag
   \end{aligned}\\
   \text{subject to: } &D_{avg}(\mathbf{M}) \leq D_{avgMax}, \label{avgmax}\\
   &M_{k}\geq m^{(l_{\min})}, \;\forall \text{ }k\in A, \label{fairnesst}\\
   &\sum ^{K}_{k=1} M_{k}\leq C/B. \label{cont}
\end{align}
\end{small} 
Constraint (\ref{avgmax}) is for the maximum average delay requirement and  (\ref{fairnesst}) is the fairness constraint for the locally cached files.  Lastly, (\ref{cont}) imposes the cache capacity constraint. Due to constraint (\ref{fairnesst}), at most $\acute{K}=min(\frac{C/B}{m^{(l_{\min})}},K)$ different files can be stored in the SBS caches. If the most popular $\acute{K}$ files are cached according to the delay constraint $D_{max}$, $m^{(l_{\min})}B$ bits allocated to each file, then the cache memory size and the fairness constraints are satisfied. If the constraint (\ref{avgmax}) is also satisfied, i.e., $D_{avgMax}=D_{max}$, then the aforementioned assignment is the  optimal and no further steps are needed. Otherwise, in order to decrease $D_{avgMax}$, the least popular file in $A$ is removed and Algorithm 1 is applied to find the optimal cache allocation for the remaining files.\\
\indent The use of Algorithm 1 ensures that the allocation yields the minimum possible average cumulative re-buffering delay for the given cache capacity constraint. Using these procedure we increase the average cost by the least possible amount while decreasing the average delay by the highest possible amount. This step is repeated until all the constraints are satisfied. The overall procedure is illustrated in Algorithm 2.
\begin{algorithm}[t]{\footnotesize
    \SetKwInOut{Input}{Input}
    \SetKwInOut{Output}{Output}
    \Input{$B$,$C$, $D_{avgMax}$}
    \Output{$\mathbf{M}$}
    $M_{k} \gets 0, k\in\left\{1,\ldots,K\right\}$\;
		$\tilde{C} \gets C/B$\;
    \For{$k \in\left\{1,\ldots,K\right\} $} {
        \If{ $\tilde{C} \geq m^{(l_{\min})} $} {
        $M_k \gets m^{(l_{\min})} $\; 
				$\tilde{C} \gets \tilde{C} - m^{(l_{\min})}$\;
        }
    }
    \textbf{execute} Algorithm 1\;
    \While { $D_{avg} > D_{avgMax}$} {
        $k=\argmin\left\{p_{i}\right\}, i\in\left\{1,\ldots,K : M_i > 0 \right\} $\;
        $M_{k} \gets 0$\;
        \textbf{execute} Algorithm 1\;
    }
    \caption{Delay constrained cost minimization}}
\end{algorithm}
\section{Numerical results}
\subsection{Simulation setup}
In this section, we evaluate the performance of the coded caching policies described in Algorithms 1 and 2. For the simulations we consider a video library of 10000 files. The popularity of the files are modeled using a Zipf distribution with parameter $w$, which adjusts its skewness. In the simulations we consider $w \in \{ 0.75, 0.85, 0.95\}$ and $T=10$. Further we set $D_{max}=10$. For the simulations we consider two different scenarios. In the first scenario we consider the cache sizes, normalized over the library size, $\hat{C}\in [0.1,0.7]$. For the given cache sizes, the maximum delay constraint $D_{max}$ can be satisfied for each video file; and hence, in the first part of simulations we analyze the average cumulative re-buffering delay. In the second case, we consider $\hat{C} = 0.08$ where the maximum delay constraint $D_{max}$ cannot be satisfied for all the files; and thus, in the second part of the simulations we analyze the trade-off between the average cost and the average cumulative re-buffereing delay. 
\subsection{Simulation results}
\begin{figure*}
    \centering
    \begin{subfigure}{0.3\textwidth}
    	\centering
        \includegraphics[scale=0.45]{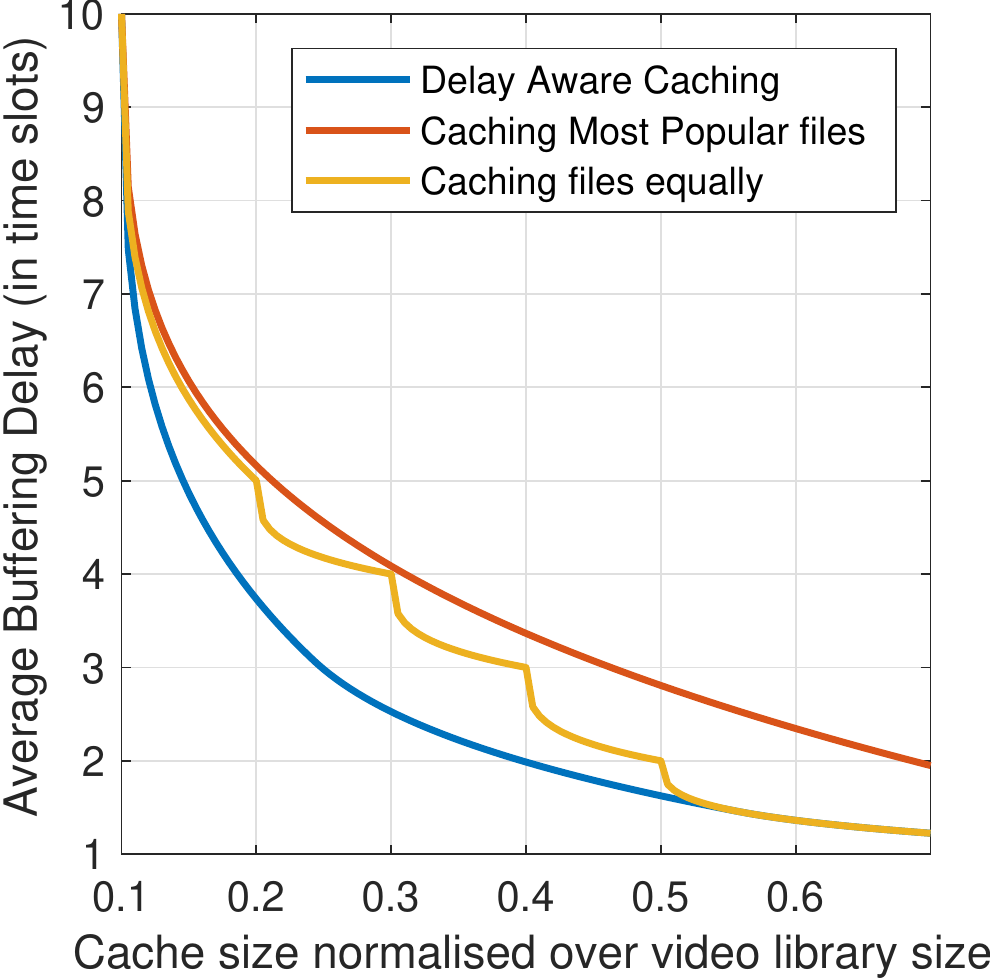}
        \caption{$w=0.75$}
    \end{subfigure}
	\begin{subfigure}{0.3\textwidth}
    	\centering
        \includegraphics[scale=0.45]{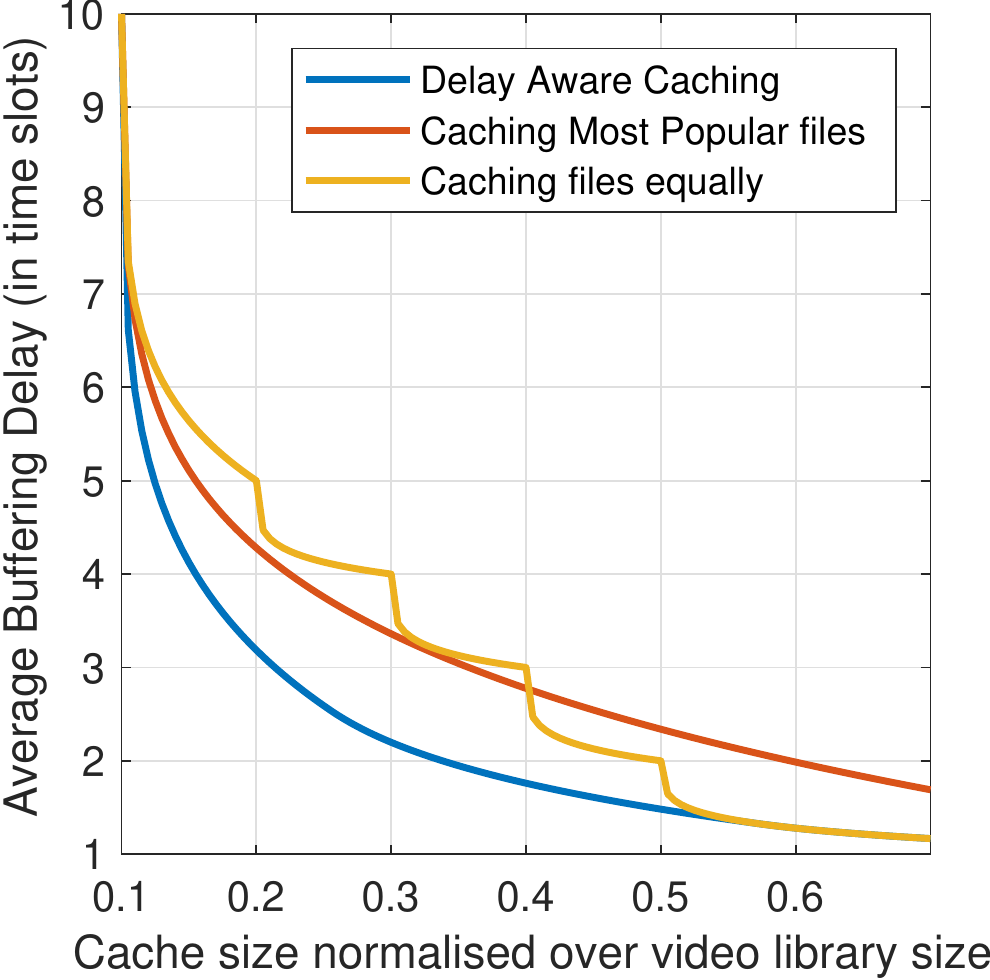}
        \caption{$w=0.85$}
    \end{subfigure}
    \begin{subfigure}{0.3\textwidth}
    	\centering
        \includegraphics[scale=0.45]{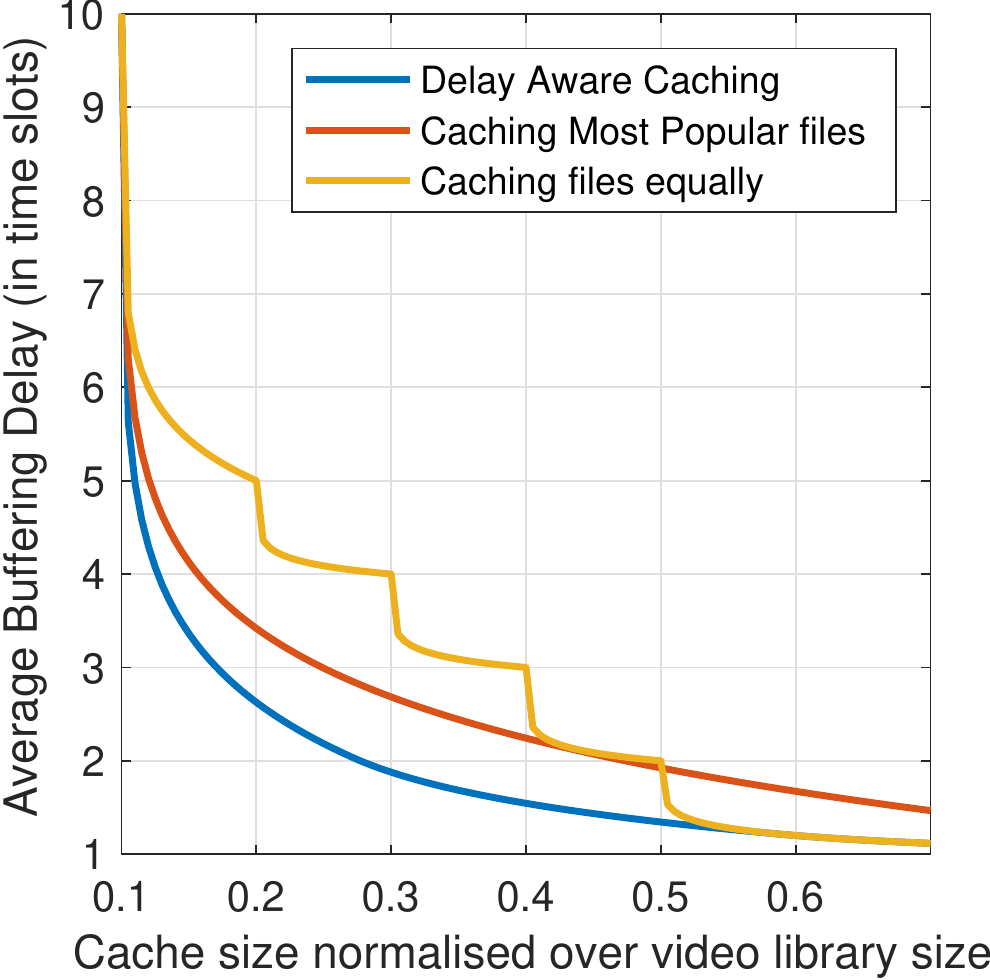}
        \caption{$w=0.95$}
    \end{subfigure}
    \caption{Cumulative buffering delay versus cache size for $D_{max}=10$ slots}
    \label{fig:mobile}
\end{figure*}

\begin{figure*}
    \centering
    \begin{subfigure}{0.3\textwidth}
    	\centering
        \includegraphics[scale=0.45]{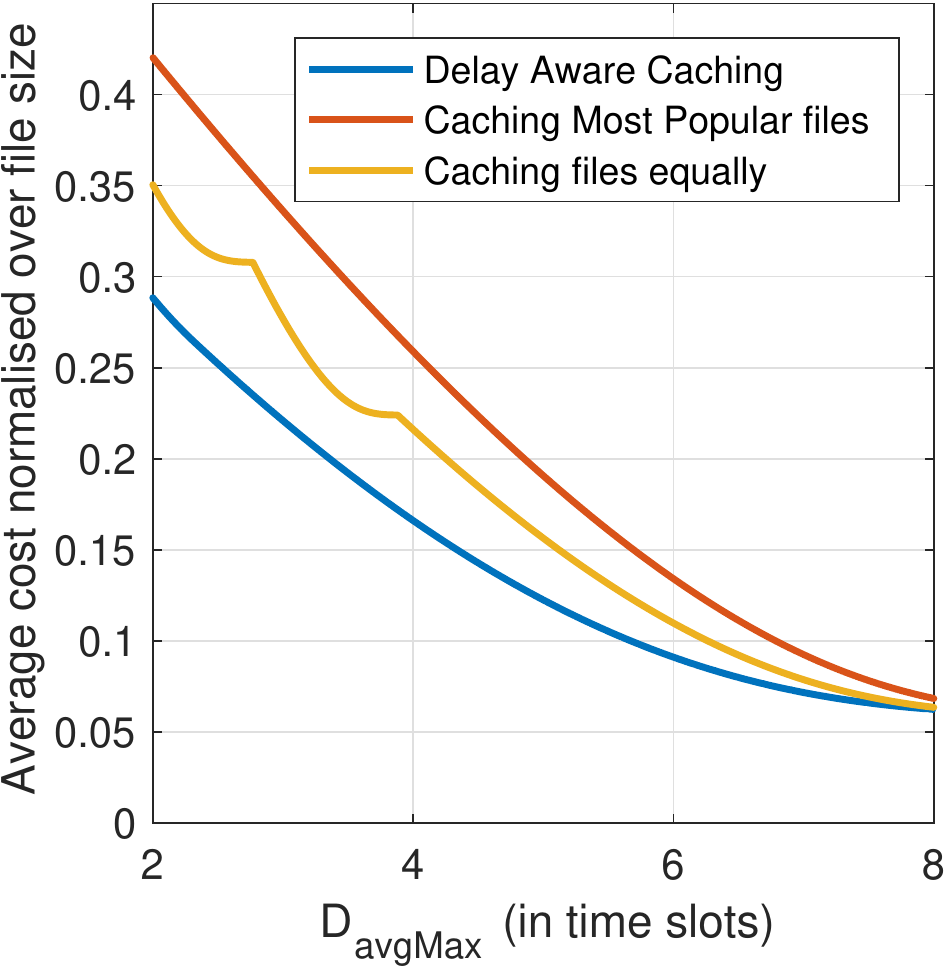}
        \caption{$w=0.75$}
    \end{subfigure}
	\begin{subfigure}{0.3\textwidth}
    	\centering
        \includegraphics[scale=0.45]{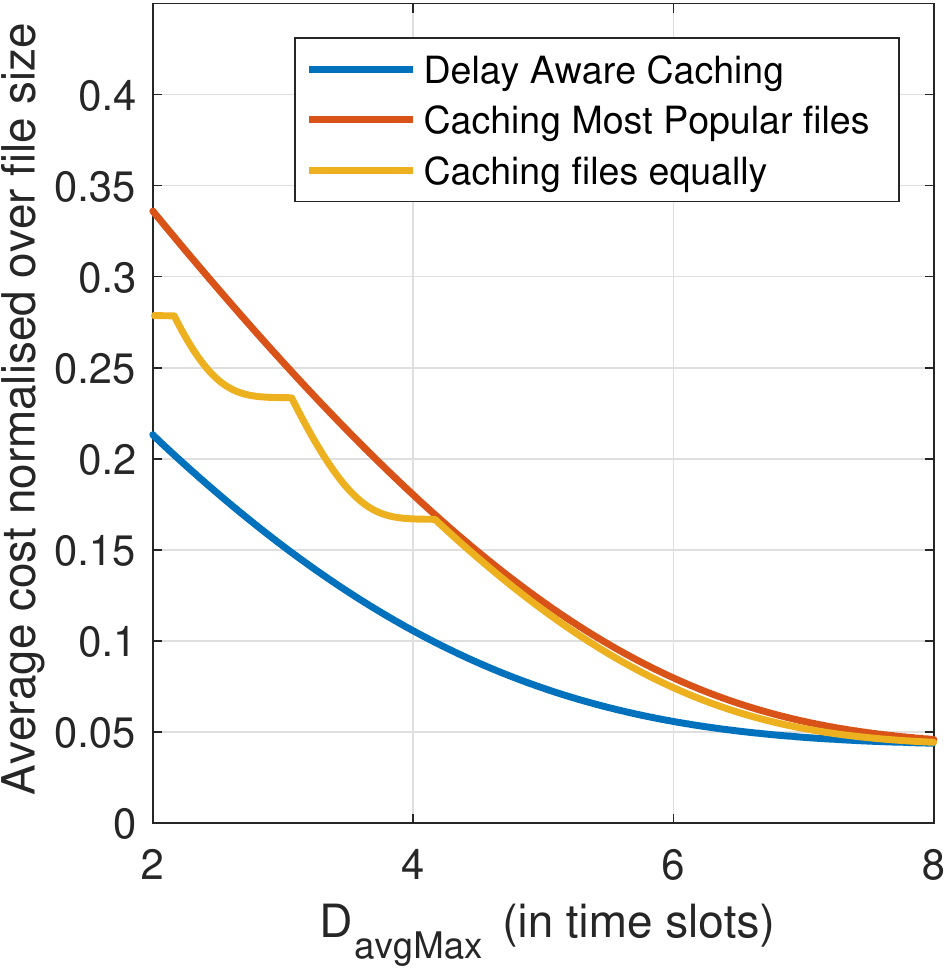}
        \caption{$w=0.85$}
    \end{subfigure}
    \begin{subfigure}{0.3\textwidth}
    	\centering
        \includegraphics[scale=0.45]{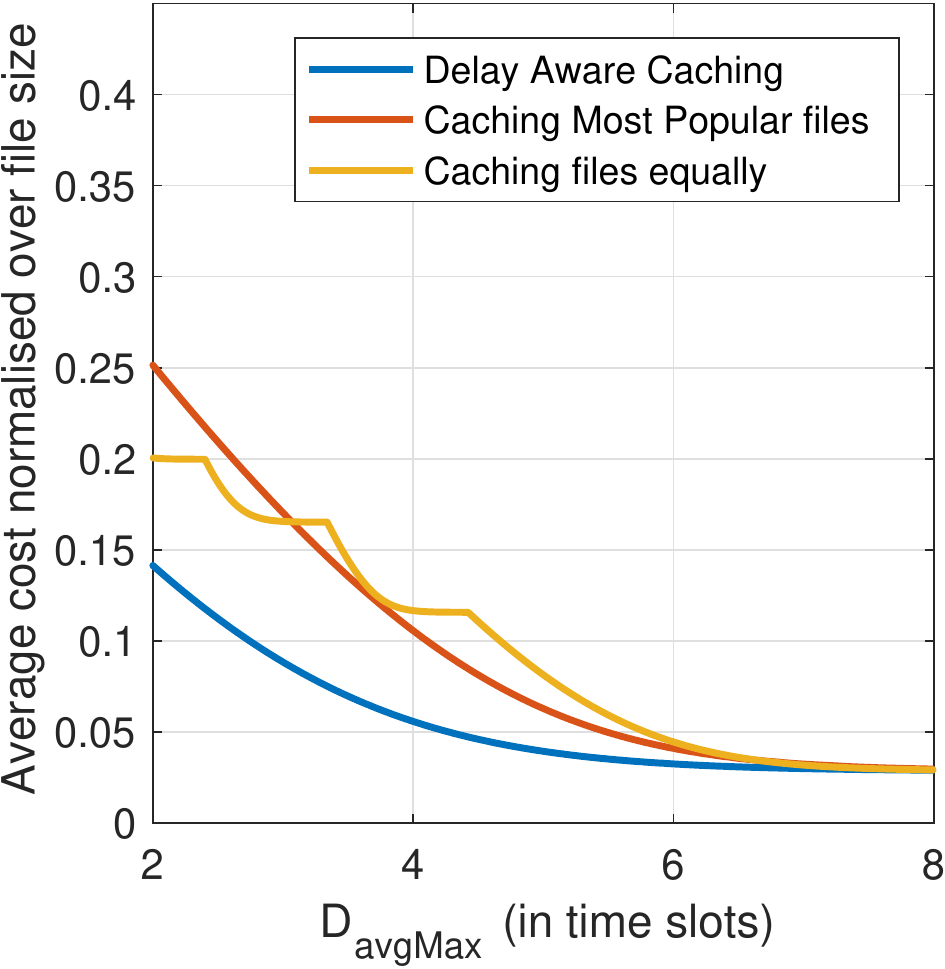}
        \caption{$w=0.95$}
    \end{subfigure}
    \caption{Average cost versus maximum average delay constraint for highly mobile users and $T = 10$ slots}
    \label{fig:mobileUser}
\end{figure*}

In the simulations we consider tow benchmarks, namely;\textit{most popular file caching} (MPFC) and the \textit{equal file caching} (EFC). In MPFC, initially, a cache size enough to satisfy $D_{max}$ is allocated to all files, then, starting from the most popular file, allocated cache size is made equal to file size until no space is left in the caches of SBSs. In EFC, again we use the same initial cache size allocation, then starting from the most popular file the allocated cache size is increased to the next decrement point. Once, the cache size of the each file is aligned to the next decrement point, we go back to the most popular file and repeat the process until no empty space is left in the caches.\\
\indent In the first simulation scenario, the cost-free delay minimization algorithm is executed and the results are shown in Figure \ref{fig:mobile}. The average cumulative re-buffering delay of the system is plotted against the available cache size for the proposed caching scheme and the two benchmarks, for three different values of $w$. The proposed caching policy is observed to have better performance than the two benchmarks in all the scenarios, and in some points the average delay is reduced up to 35\% with respect to the benchmark with the best performance at this point. From Figure \ref{fig:mobile}, it is also clear that for highly skewed distributions (libraries with a few very popular videos), MPFC performs closer to the proposed algorithm, while for less skewed distributions, the second benchmark is closer.\\ 
%The performance of EFC exhibits steps that occur every time all the files have been cached at the same delay level $D^{(l)}$. 
In the second simulation scenario, in which the cache size is not sufficient to satisfy delay constraint $D_{avgMax}$, Algorithm 2 is executed, and its performance  is compared with MPFC and EFC policies in Figure \ref{fig:mobileUser}.\\ 
\indent MPFC with a given $D_{avgMax}$ constraint is executed according to the following strategy: first the most popular $\frac{C/B}{m^{(l_{\min})}}$ files are cached according to the maximum allowed delay $D_{Max}$. If the average delay constraint is not satisfied, i.e., $D_{max}>D_{avgMax}$ then the least popular file that is cached is removed, and the corresponding cache memory is used for the most popular file that is not cached up to the maximum level. This procedure is repeated until the average delay constraint $D_{avgMax}$ is satisfied for all the cached files. For the EFC benchmark, again after the initial step, if the average delay constraint $D_{avgMax}$ is not satisfied, then the least popular file in the cache is removed. The equal file caching algorithm described above is  applied subsequently on the files that are still in the cache. This procedure is repeated until the average delay constraint is satisfied for all the cached files. The three plots portray the relationship between the average cost and the  average delay constraint $D_{avgMax}$. Our proposed solution exhibits significant improvement in comparison with the benchmark policies. For example, for $w=0.95$ and $D_{avgMax}=2$, the average cost is improved by 30\% and 44\% with respect to EFC and MPFC, respectively. As it is expected, the tighter the  average delay constraint $D_{avgMax}$, the higher the cost. 
%Also, as the constraint increases, the average cost asymptotically tends to its minimum achievable value, which is the sum of probabilities of video files that are not stored at the SBS caches, if $m^{l_{min}}=1$ fragment is cached for each of the $\frac{C/B}{m^{(l_{\min})}}=\frac{\hat{C}KT}{m^{(l_{\min})}}=8000$ most popular files. 
Lastly, for all the three caching policies the cost decreases as the skewness coefficient $w$ increases. This is attributed to the fact that the  popularity of less popular files is lower for more skewed distributions.

\section{Conclusion}
We studied the cache capacity-delay trade-off in heterogeneous networks with a focus on continuous video display targeting streaming applications. We first proposed a caching policy that minimizes the average cumulative re-buffering delay under the high mobility assumption. We then considered a scenario in which the average cumulative re-buffering delay is a given system requirement, and introduced a caching policy that minimizes the amount of data downloaded from the MBS while satisfying this requirement. Numerical simulations have been presented, showcasing the improved performance of the proposed caching policy in comparison with other benchmark caching policies. General user mobility patterns will be studied as a future extension of this work.




%\bibliographystyle{IEEEtran}
\bibliographystyle{IEEEtran}
\bibliography{IEEEabrv,conf}
\end{document}